\documentclass[a4paper,12pt]{article}
\usepackage[pctex32]{graphicx}

\newcommand{\be}{\begin{equation}}
\newcommand{\ee}{\end{equation}}
\newcommand{\ben}{\begin{eqnarray}\displaystyle}
\newcommand{\een}{\end{eqnarray}}

\makeatletter
\newcommand{\figcaption}{\def\@captype{figure}\caption}
\newcommand{\tabcaption}{\def\@captype{table}\caption}
\makeatother

\begin{titlepage}

\begin{document}
{\baselineskip20pt


\vskip .6cm

\begin{center}
{\Large \bf Thermodynamic Geometry of the Born-Infeld-anti-de Sitter
black holes}

\end{center} }

\vskip .6cm
 \centerline{\large Peng Chen}

\vskip .6cm

\begin{center}
{Institute of Theoretical Physics\\
Chinese Academy of Sciences, Beijing 100190, PRC}
\end{center}

\vspace{5mm}


\begin{abstract}
Thermodynamic geometry is applied to the Born-Infeld-anti-de Sitter
black hole (BIAdS) in the four dimensions, which is a nonlinear
generalization of the Reissner-Norstr\"om-AdS black hole (RNAdS). We
compute the Weinhold as well as the Ruppeiner scalar curvature and
find that the singular points are not the same with the ones
obtained using the heat capacity. Legendre-invariant metric proposed
by Quevedo and the metric obtained by using the free energy as the
thermodynamic potential are obtained and the corresponding scalar
curvatures
 diverge at the Davies points.
\end{abstract}


\vskip 0.8cm

\noindent chenpeng@itp.ac.cn \\

\noindent
\end{titlepage}

\setcounter{page}{2}

\section{Introduction}

Born-Infeld electrodynamics was first introduced by Born and Infeld
in 1930's to remove the divergence of the electron's
self-energy~\cite{BI}. It has received renewed attention for the
last fifteen years since it arises naturally in open superstrings
and in D-branes~\cite{FTL,Tse,Lei}. On the other hand,
thermodynamics of black holes in AdS space has also generated
renewed attention due to the AdS/CFT duality~\cite{adsref}, which
relates thermodynamics of black holes in AdS space to the dual CFT
in a lower dimension. Studying the phase transitions of AdS black
holes is an effective way of exploring phase structures in the dual
field theories. Born-Infeld-AdS black hole solution and the
thermodynamic properties has been analyzed in~\cite{Dey}-\cite{mkp}.

Introducing differential geometric concepts into ordinary
thermodynamics was first done by Weinhold \cite{wein}. He proposed a
Riemannian metric defined as the second derivatives of internal
energy $U$ with respect to entropy and other extensive quantities of
a thermodynamic system. However, the geometry of this metric seems
to have no physical meaning in the context of equilibrium
thermodynamics. Few years later, Ruppeiner \cite{rupp} introduced
another metric, defined as the negative Hessian of entropy $S$ with
respect to the internal energy and other extensive quantities of a
thermodynamic system. In \cite{ruppwein} it was proved that the
Ruppeiner metric is conformal to the Weinhold metric with the
inverse temperature as the conformal factor. The Ruppeiner geometry
has its physical meaning in the fluctuation theory of equilibrium
thermodynamics. Both metrics have been applied to study the geometry
of the thermodynamics of ordinary systems~\cite{ord}-\cite{jjka}. In
particular, it was found that the Ruppeiner geometry carries
information of phase structure of thermodynamic system; and scalar
curvature of the metric diverges at the phase transition and
critical point, which shows interaction of the system. For
thermodynamic systems with no statistical mechanical interactions
(for example, ideal gas), the scalar curvature is zero and the
Ruppeiner metric is flat. Because of the success of their
applications to ordinary thermodynamic systems, they have also been
used to study black hole phase structures and lots of results have
been obtained for
different sorts of black holes~\cite{fgk}-\cite{wei10}.\\
~~However, the results they present are sometimes contradictory. For
instance, for the RN black hole: the Weinhold metric predicts phase
transitions which are compatible with standard black hole
thermodynamics, while the Ruppeiner curvature is flat, giving no
information at all about phase transition.
 To overcome this
inconsistency, the theory of Gometrothermodynamics(GTD) was proposed
recently~\cite{quev07,quev08,quev10}. It incorporates arbitrary
Legendre transformations into the geometric structure of the
equilibrium space in an invariant manner. In~\cite{lvhong} other
thermodynamic potentials were proposed and metrics on all these
thermodynamic potentials were investigated. They showed that in
general for a system with $n$-pairs of intensive/extensive
variables, all thermodynamical potential metrics can be embedded
into a flat $(n,n)$-dimensional space. In this paper we will apply
thermodynamic geometry into BIAdS black holes and see whether the
information about phase transition represented by divergence of heat
capacity can be reproduced.

This paper is organized as follows. In Section II we review the
basics of Born-Infeld-AdS black hole and its most important
thermodynamic quantities, plot the graphics of the heat capacities
for fixed charge and fixed potential. In Section III we apply
thermodynamic geometry to BIAdS black holes, getting the metrics and
their corresponding scalar curvatures. In the last section we
discuss our results and suggestions.

\section{Basics of BIAdS black holes}

First let's consider the BIAdS action which is a (3+1)-dimensional
gravity coupled with nonlinear electrodynamics ~\cite{Dey,fk,cpw}
\begin{equation}\label{action}
S=\int d^4x \sqrt{-g}\left(\frac{R-2\Lambda}{16\pi G}+L(F)\right),
\end{equation}
where
\begin{equation}\label{actionF}
L(F)=\frac{b^2}{4\pi G}\left(1-\sqrt{1+\frac{2F}{b^2}}\right)
\end{equation}
with $F\equiv F^{\mu\nu}F_{\mu\nu}/4$, where the constant $b$ is the
Born-Infeld parameter, and $\Lambda = - 3/l^2$ is the cosmological
constant.  Note that this Lagrangian reduces to the RNAdS one in the
limit $b^2 \rightarrow \infty$.

After solving the equations of motion for the gauge field $A_\mu$
and the gravitational field $g_{\mu\nu}$, the BIAdS black hole
solutions~\cite{Dey,fk,cpw} can be written as
\begin{equation} \label{metric}
ds^2 = - f(r) dt^2 + f(r)^{-1} dr^2 + r^2  d \Omega^2.
\end{equation}
Here, the metric function $f(r)$ is given by
\begin{eqnarray}\label{metric f}
f(r) &=& 1 - \frac{2M}{r} + \frac{r^2}{l^2} + \frac{2 b^2 r^2}{3} \left( 1 - \sqrt{ 1 + \frac{Q^2}{b^2r^4 }} \right)  \nonumber \\
&{}& +~\frac{ 4 Q^2}{ 3 r^2} \hspace{0.2cm}  {\cal F} \left(
\frac{1}{4}, \frac{1}{2}, \frac{5}{4}, -\frac{Q^2}{b^2 r^4 }
\right),
\end{eqnarray}
where ${\cal F}$ is a hypergeometric function. The only non-zero
component with the electric charge $Q$ is given by $F^{01} = -E  = -
Q/{\sqrt{ r^4 + {Q^2}/{b^2}}}$. Hereafter we only consider $Q\ge 0$
and $b\ge0$ without any loss of generality. In the limit $Q
\rightarrow 0$, $f(r)$ reduces to the Schwarzschild-anti de Sitter
black hole (SAdS) case, while in the limit $b \rightarrow \infty$
and $Q \neq 0$, $f(r)$ reduces to the RNAdS case.

We can solve the equation $f(r) = 0$ to get the ADM mass $M$ which
is given by
\begin{eqnarray}
\label{mass} M(r_+,Q,b) &=& \frac{r_{+}}{2} + \frac{r_{+}^3}{2l^2} +
\frac{ b^2 r_{+}^3}{3}
\left( 1 - \sqrt{ 1 + \frac{Q^2}{ b^2 r_+^4}} \right) \nonumber \\
&{}& +~\frac{ 2 Q^2}{ 3 r_+} \hspace{0.2cm}  {\cal F} \left(
\frac{1}{4}, \frac{1}{2}, \frac{5}{4}, -\frac{Q^2}{ b^2 r_+^4}
\right)
\end{eqnarray}
with the outer horizon $r=r_+$. If we demand both $f(r)$ and
${df(r)}/{dr}$ vanish at the degenerate horizon, then we get the
extremal BIAdS. The radius $r^2_{e}$ of the extremal BIAdS  is given
by
\begin{equation}
r_{e}^2 = \frac{l^2}{6}\left(\frac{1 + \frac{3}{2b^2 l^2}}{1 +
\frac{4}{4b^2 l^2}}\right) \left[- 1 + \sqrt{ 1 + \frac{12\left(1
+ \frac{3}{4b^2 l^2}\right)}{b^2l^2\left(1 + \frac{3}{2b^2 l^2}
\right)^2} \left(b^2Q^2 -\frac{1}{4} \right)}~\right].
\end{equation}
So the condition $ b Q \ge 0.5$ must be satisfied in order to have a
real root for $r^2_{e}$. As a result, the parameter space for the
BIAdS is
\begin{equation}
0.5 \leq bQ \leq \infty.
\end{equation}
We call the lower bound ($bQ=0.5$)as the critical BIAdS and the
upper bound of $b \to \infty$ as the RNAdS.

The Hawking temperature $T_H=f'(r_+)/4\pi$ is given by
\begin{equation}
\label{temp} T_H(r_+,Q,b) = \frac{1}{4\pi} \left( \frac{1}{r_+} +
\frac{3r_+}{l^2} + 2 b^2 r_+ \left[1 - \sqrt{1 + \frac{Q^2}{b^2
r_+^4}}~\right]  \right).
\end{equation}
Note that in the limit $Q \rightarrow 0$,  $T_H$ reduces to the SAdS
case, while in the limit of $b \rightarrow \infty $ and $Q \neq 0$,
$T_H$ reduces to the RNAdS case.\\

 Then, using Eqs. (\ref{mass})
and (\ref{temp}), the heat capacity $C(r_+,Q,b)=(dM/dT_{H})_Q$ for
fixed-charge is
\begin{equation}\label{cq}
C_{Q} = \frac{2 \pi r^2_+ \sqrt{1 + \frac{Q^2}{b^2 r^4_+}} \left[ 3
r^4_+ + l^2 \left(r^2_+ + 2 b^2 \left(1 - \sqrt{1 + \frac{Q^2}{b^2
r^4_+}}\right) r^4_+\right) \right]} { 3 \sqrt{1 + \frac{Q^2}{b^2
r^4_+}} r^4_+ + l^2 \left[- \sqrt{1 + \frac{Q^2}{b^2 r^4_+}} r^2_{+}
+2 Q^2  - 2 b^2 \left( 1 - \sqrt{1 + \frac{Q^2}{b^2 r^4_+}} \right)
r^4_+ \right]}. \end{equation} Note that  in the limit $Q
\rightarrow 0$, the heat capacity $C_{Q}$ reduces to the SAdS case
as
\begin{eqnarray}\label{aaSc}
C_{Q}^{SAdS}(r_+)&=& 2\pi r_+^2 \Big(\frac{3r_+^2+l^2}{3r_+^2 - l^2
}\Big), \label{aaSf}
\end{eqnarray}
In the limit of $b^2 \rightarrow \infty$, $C_{Q}$ reduces to the
RNAdS case as
\begin{eqnarray}\label{aac2}
C_{Q}^{RNAdS}(r_+,Q)&=& 2\pi r_+^2
\Big[\frac{3r_+^4+l^2(r_+^2-Q^2)}{3r_+^4+l^2(-r_+^2+3Q^2)}\Big],
\label{aaf2}
\end{eqnarray} According to the well-known
Area-Entropy formula, the Bekenstein-Hawking entropy of the BIAdS
black hole is
\begin{equation}\label{entropy}
S=\pi r_{+}^2.
\end{equation}
By the first law of thermodynamics,
\begin{equation}\label{1st law}
dM=TdS+\Phi dQ
\end{equation}
we can get the electric potential as
\begin{equation}
\Phi = \frac{Q}{r_+}  {\cal F} \left( \frac{1}{4}, \frac{1}{2},
\frac{5}{4}, -\frac{Q^2}{ b^2 r_+^4} \right)
\end{equation}
The heat capacity for fixed potential is
\begin{equation}\label{cfi}
C_{\Phi}=-\frac{2 \pi  r_{+} \left(r_{+} \left(\frac{3}{l^2}-2 b^2
\left(\sqrt{\frac{Q^2}{b^2
r_{+}^4}+1}-1\right)\right)+\frac{1}{r_{+}}\right)}{\frac{4
   Q^2}{r_{+}^4 \sqrt{\frac{Q^2}{b^2 r_{+}^4}+1} \left(\sqrt{\frac{Q^2}{b^2 r_{+}^4}+1}{\cal F} \left( \frac{1}{4}, \frac{1}{2},\frac{5}{4},-\frac{Q^2}{b^2
   r_{+}^4}\right)+1\right)}+2 b^2 \left(\sqrt{\frac{Q^2}{b^2 r_{+}^4}+1}-1\right)-\frac{4 Q^2}{r_{+}^4 \sqrt{\frac{Q^2}{b^2
   r_{+}^4}+1}}-\frac{3}{l^2}+\frac{1}{r_{+}^2}}\end{equation}

Here we plot $C_{Q}$ and $C_{\Phi}$ as follows

   \begin{center}
   \includegraphics[width=10cm]{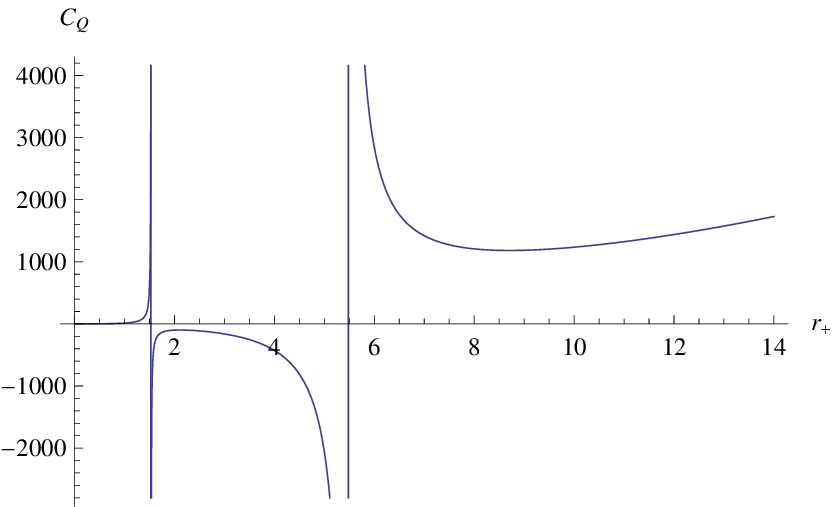}
\figcaption{The heat capacity for fixed charge $C_{Q}$ vs $r_{+}$
with $b=0.5,l=10,Q=1$. At $r_{+}=1.5307,  5.4781$, $C_{Q}$
diverges.} \label{fig1}
\end{center}

\begin{center}
     \includegraphics[width=14cm]{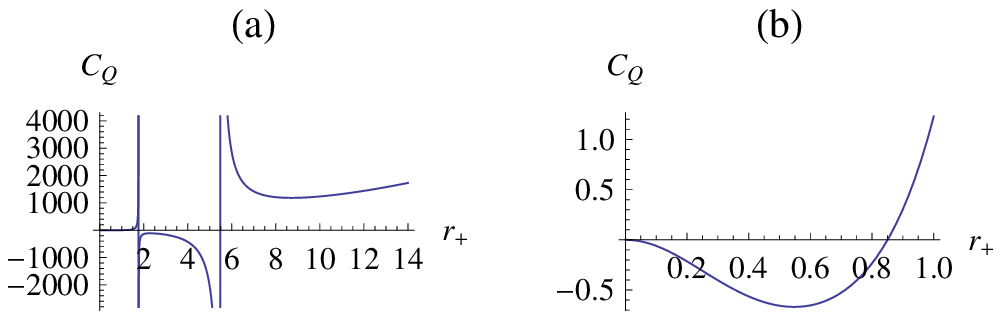}
\figcaption{The heat capacity for fixed charge $C_{Q}$ vs $r_{+}$
with $b=1,l=10,Q=1$.(a)At $r_{+}=1.7685/6,  5.4774/5$, $C_{Q}$
diverges.(b)At $r_{+}=0.850492$, $C_{Q}$ vanishes.} \label{fig2}
\end{center}

\begin{center}
   \includegraphics[width=10cm]{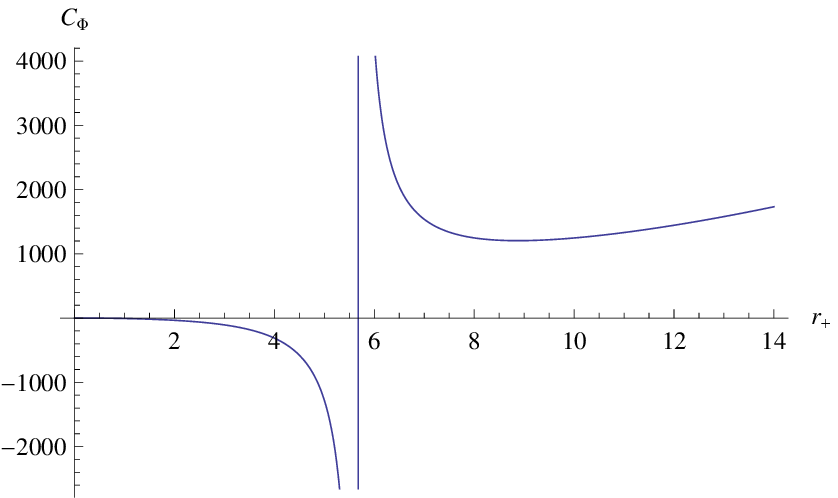}
\figcaption{The heat capacity for fixed potential $C_{\Phi}$ vs
$r_{+}$ with $b=0.5,l=10,Q=1$. At $r_{+}=5.6835/6$, $C_{\Phi}$
diverges.} \label{fig3}
\end{center}

\begin{center}
   \includegraphics[width=14cm]{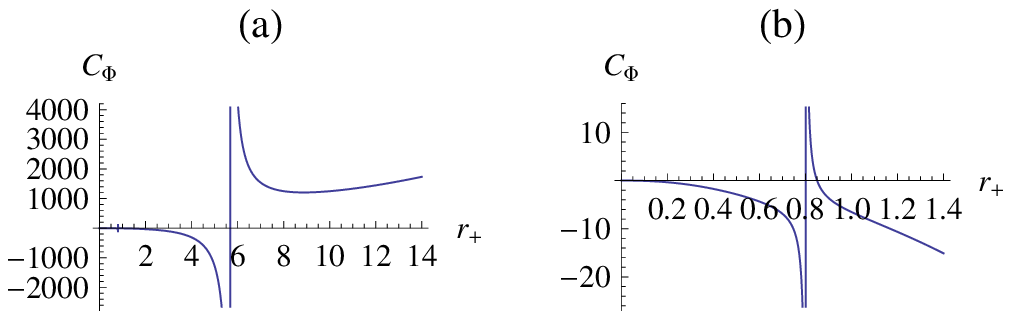}
\figcaption{The heat capacity for fixed potential $C_{\Phi}$ vs
$r_{+}$ with $b=1,l=10,Q=1$. (a) At $r_{+}=5.6834/5$, $C_{\Phi}$
diverges. (b)$C_{\Phi}$ diverges at 0.8009/11, vanishes at
0.8504/6.} \label{fig4}
\end{center}

We may also introduce analogously charge capacitances at fixed
temperature or entropy, they are given by
\begin{equation}\label{cts}
\widetilde{C}_{T}\equiv\frac{\partial Q}{\partial
\Phi}|_{T}=\frac{\frac{\partial Q}{\partial
r_{+}}|_{T}}{\frac{\partial \Phi}{\partial r_{+}}|_{T}},
\widetilde{C}_{S}\equiv\frac{\partial Q}{\partial
\Phi}|_{S}=\frac{\frac{\partial Q}{\partial
r_{+}}|_{S}}{\frac{\partial \Phi}{\partial r_{+}}|_{S}}
\end{equation}
It turns out that $\widetilde{C}_{S}$ is positive definite(for
$b=0.5,1;l=10,Q=1$), and there are no sigular points associated with
it, so we will skip it in the following discussion.
$\widetilde{C}_{T}$ is plotted as follows.

\begin{center}
   \includegraphics[width=10cm]{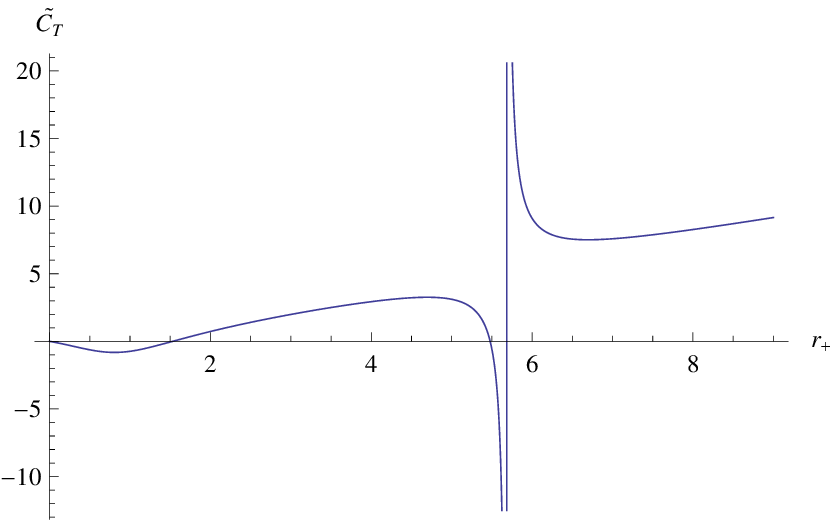}
\figcaption{The charge capacitance at fixed temperature
$\widetilde{C}_{T}$ vs $r_{+}$ with $b=0.5,l=10,Q=1$. At
$r_{+}=5.6835/6$, $\widetilde{C}_{T}$ diverges. $\widetilde{C}_{T}$
vanishes at 1.5307, 5.4781.} \label{fig5}
\end{center}

\begin{center}
   \includegraphics[width=10cm]{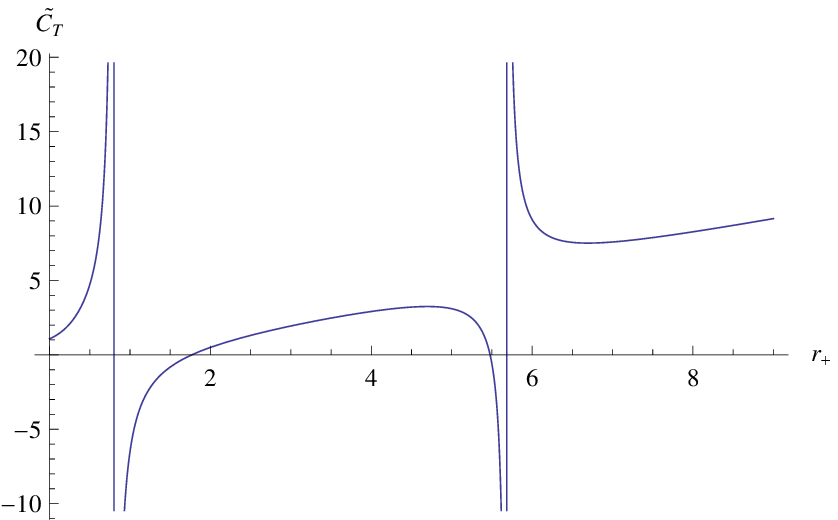}
\figcaption{The charge capacitance at fixed temperature
$\widetilde{C}_{T}$ vs $r_{+}$ with $b=1,l=10,Q=1$. At
$r_{+}=0.8009/11,  5.6834/5$, $\widetilde{C}_{T}$ diverges.
$\widetilde{C}_{T}$ vanishes at 1.7685/6, 5.4774/5.} \label{fig6}
\end{center}

From Figure 1 and Figure 5 (both are for $b=0.5$), we see clearly
that when $C_{Q}$ diverges(at $r_{+}=1.5307, 5.4781$),
$\widetilde{C}_{T}$ vanishes. From Figure 2 and Figure 6 (both are
for $b=1$), we see clearly that when $C_{Q}$ diverges (at
$r_{+}=1.7685/6,  5.4774/5$), $\widetilde{C}_{T}$ vanishes. It is
consistent with the observation in~\cite{lvhong} that the vanishing
of a certain capacity is always associated with the divergence of
another.

\section{Thermodynamic geometry of BIAdS black holes}
Study of black hole phase transitions from the point of view of
thermodynamic geometry has been quite active recently. In this
section, we will apply thermodynamic geometry to BIAdS black holes.
\subsection{The Weinhold metric}
The Weinhold metric is defined as \cite{wein}

\begin{equation}
dS_{W}^2=g_{ij}^{W}dX^i dX^j \label{wein}
\end{equation}
where
\begin{equation}
g_{ij}^{W}=\frac{\partial^2{M(X^k)}}{\partial X^i\partial
X^j},~~{\textrm{and}}~~X^i\equiv X^i(S,N^a) \label{weinmet}
\end{equation}
Here $N^a$'s are all other extensive variables of the system. In
this case, $N^a$ would be the charge $Q$. We can substitute
$r_{+}=\sqrt{S}/\sqrt{\pi}$ into (\ref{mass}) and get the mass as a
function of $S$ and $Q$ (Note that in this paper $l$ and $b$ will be
fixed at some value). At this stage $dS_{W}^2$ can be calculated
straightforward. As the metric and corresponding scalar curvature
are very complicated, we will not present them here. Instead we plot
the scalar curvature as follows.
\begin{center}
   \includegraphics[width=10cm]{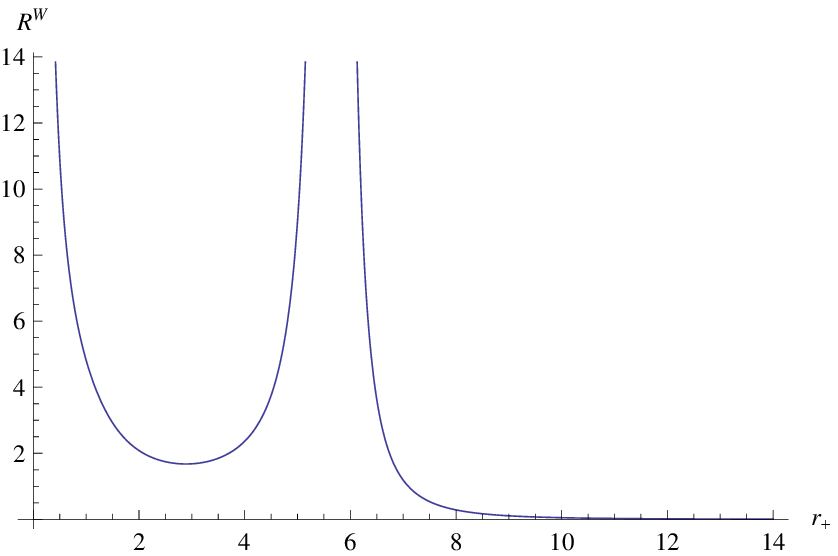}
\figcaption{The Weinhold curvature $R^{W}$ vs $r_{+}$ with
$b=0.5,l=10,Q=1$. At $r_{+}=0, 5.6835/6$, $R^{W}$ diverges.}
\label{fig7}
\end{center}

\begin{center}
   \includegraphics[width=10cm]{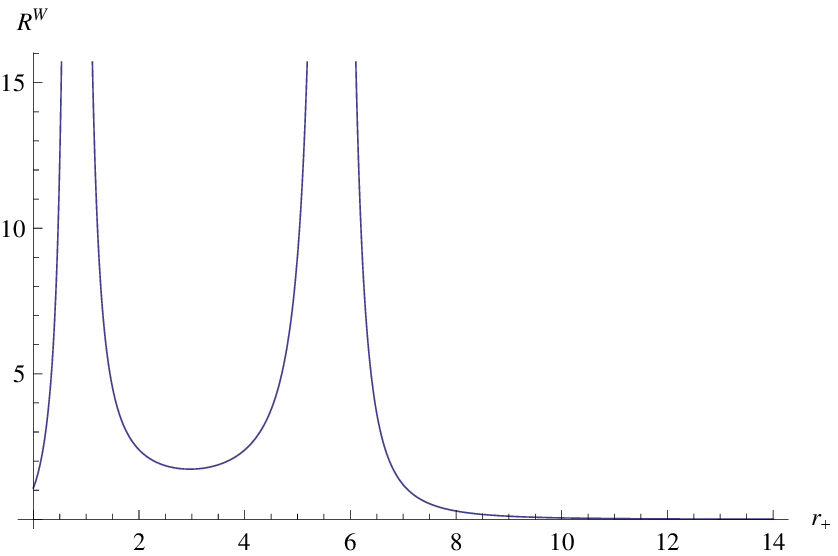}
\figcaption{The Weinhold curvature $R^{W}$ vs $r_{+}$ with
$b=1,l=10,Q=1$. At $r_{+}=0.8010/11, 5.6834/5$, $R^{W}$ diverges.}
\label{fig8}
\end{center}
For BIAdS balck holes, the Weinhold geometry is curved, signaling
interaction for this thermodynamic system. There are singular
points, but they are not consistent with the ones of the heat
capacity for fixed charge.

\subsection{The Ruppeiner metric}
The Ruppeiner metric is given by~\cite{rupp}
\begin{equation}
dS_{R}^2=g_{ij}^{R}dX^idX^j \label{rupp}
\end{equation}
where,
\begin{equation}
g_{ij}^{R}=-\frac{\partial^2{S(X^k)}}{\partial X^i\partial
X^j},~~{\textrm{and}}~~X^i\equiv X^i(M,N^a) \label{ruppmet}
\end{equation}
 For
BIAdS black hole $N^a=Q$.\\ In order to find $g_{ij}^{R}$ it is
desirable to express $S$ in terms of $M$ and $Q$. However from
(\ref{mass}) we see that M is expressed as a function of
$r_{+}(S=\pi r_{+}^2)$ and $Q$, which is a hypergeomeric function
and is invertible. Fortunately, it was proved that Ruppeiner metric
and Weinhold metric are related with each other by a conformal
factor \cite{ruppwein}
\begin{equation}
dS_{R}^2=\frac{1}{T}dS_{W}^2 \label{confrel}
\end{equation}
where $T$ is the temperature of the system. In this case this would
correspond to the Hawking temperature of the BIAdS black hole. So we
can use the Weinhold metric and the relation of the two metrics to
get the Ruppeiner metric. We plot the Ruppeiner scalar curvature as
follows.
\begin{center}
   \includegraphics[width=14cm]{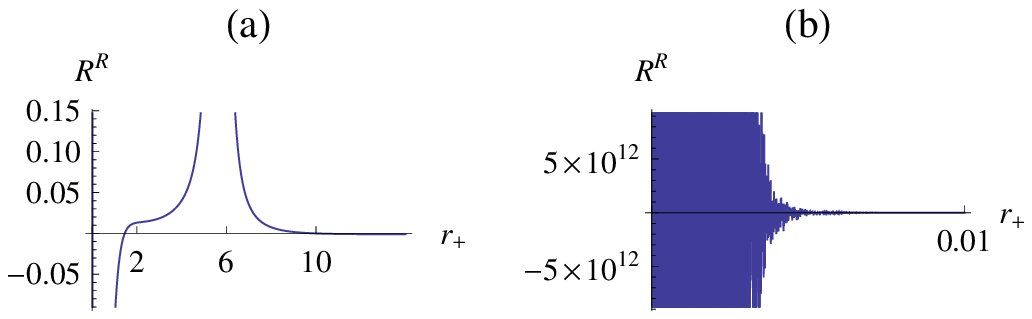}
\figcaption{The Ruppeiner curvature $R^{R}$ vs $r_{+}$ with
$b=0.5,l=10,Q=1$. At~$r_{+}=0,5.6835/6$, $R^{R}$ diverges. At
~$r_{+}$=1.4590/2, it vanishes.} \label{fig9}
\end{center}

\begin{center}
   \includegraphics[width=14cm]{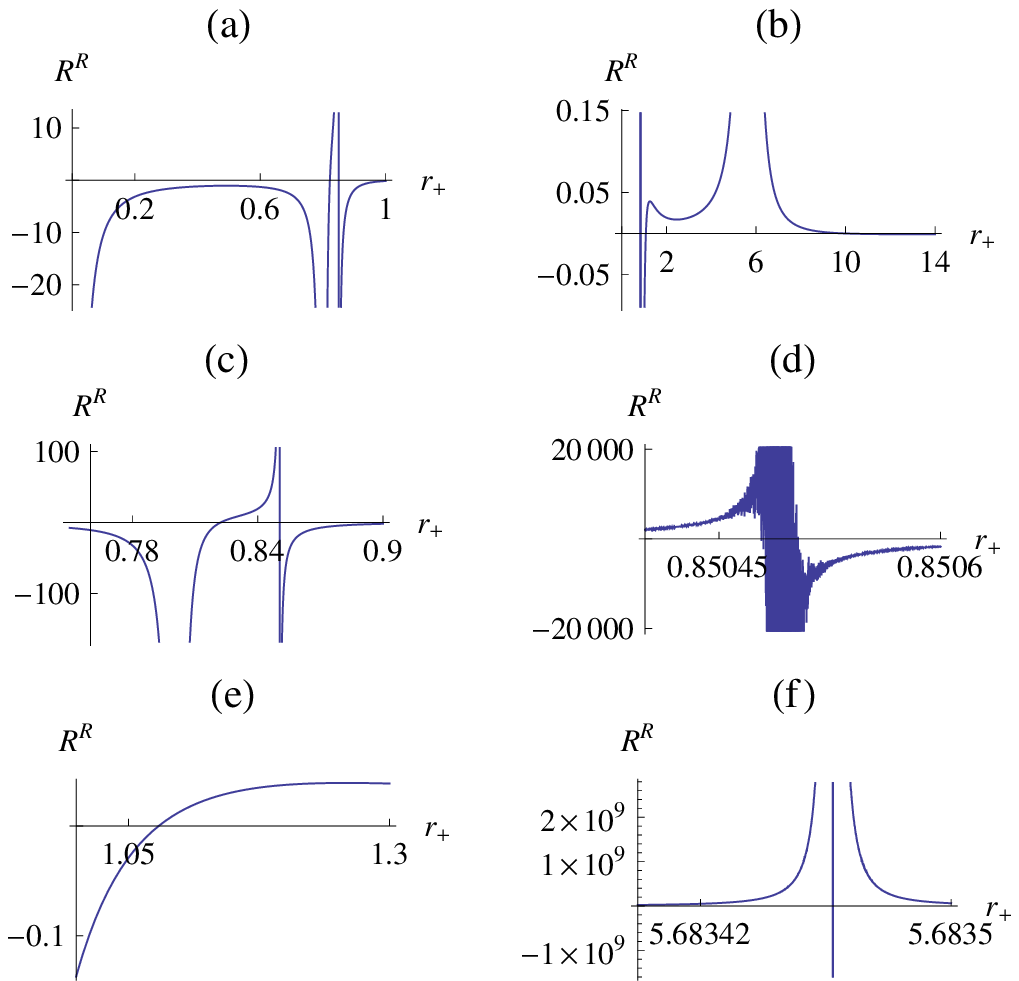}
\figcaption{The Ruppeiner curvature $R^{R}$ vs $r_{+}$ with
$b=1,l=10,Q=1$. (a)(c)(d)(f): it diverges at $r_{+}=0,  0.8010/1,
0.8504/6,  5.6834/5$ respectively;(c)(e): it vanishes at
$r_{+}$=0.8222/4,  1.0790/1} \label{fig10}
\end{center}

Like the Weinhold geometry, the Ruppeiner geometry is also curved,
signaling interaction for this thermodynamic system. There are
singular points, but they are not consistent with the ones of the
heat capacity for fixed charge(except one singular point at
$r_{+}=0.8504/6$, at which $C_{Q}$ vanishes.).

Note that for $b=0.5$, the Weinhold metric, the Ruppeiner metric all
diverge at the same points ($r_{+}=0, 5.6835/6$), but they are
inconsistent with the result of heat capacity for fixed charge
($r_{+}=1.5307,  5.4781$); for $b=1$, $C_{\Phi}$, the Weinhold
metric, the Ruppeiner metric all diverge at the same points
($r_{+}=0.8010/1,  5.6834/5$), but they are different from the
divergencies of heat capacity for fixed charge ($r_{+}=1.7685/6,
5.4774/5$). It is consistent with the observation in \cite{rupp2}
that whilst the Ruppeiner and Weinhold metrics indeed reveal the
signals of black hole phase transitions associated with divergence
of heat capacity with fixed electric potential or angular velocity,
they are insensitive to the Davies curve \cite{daviescurve} where
the heat capacity with fixed charge and/or angular momentum
diverges.

\subsection{The Quevedo metric}
Weinhold and Ruppeiner metrics are supposed to give a direct
relationship between curvature singularities and divergencies of the
heat capacity. Unfortunately, the singular points they present are
not consistent with the ones of the heat capacity. In fact, the
results they present are sometimes contradictory. For instance, for
the RN black hole: the Weinhold metric predicts phase transitions
which are compatible with standard black hole thermodynamics, while
the Ruppeiner curvature is flat, giving no information at all about
phase transition. To solve this problem, in~\cite{Shen} a
generalized Ruppeiner metric was proposed with variables $(M,Q)$
replaced by $(M-\Phi Q,\Phi)$. After the replacement, the new metric
gives the correct singular point. Another approach to solve the
puzzle was given by Quevedo, who proposed a new metric, which is
Legendre-invariant. The Legendre-invariant metric reproduces the
corresponding phase transition structure. There are many
Legendre-invariant metrics that we can
use, in this case, we will use this one:\\

\begin {equation}
g^{Q}_{ab} = (SM_S + Q M_Q)\left(
\begin{array}{cc}
-M_{SS}& 0  \\
0 &  M_{QQ}
\end{array}
\right)
 \ .
\label{grn} \end{equation} The corresponding metric and curvature
are computed and the curvature is plotted as follows.

\begin{center}
   \includegraphics[width=10cm]{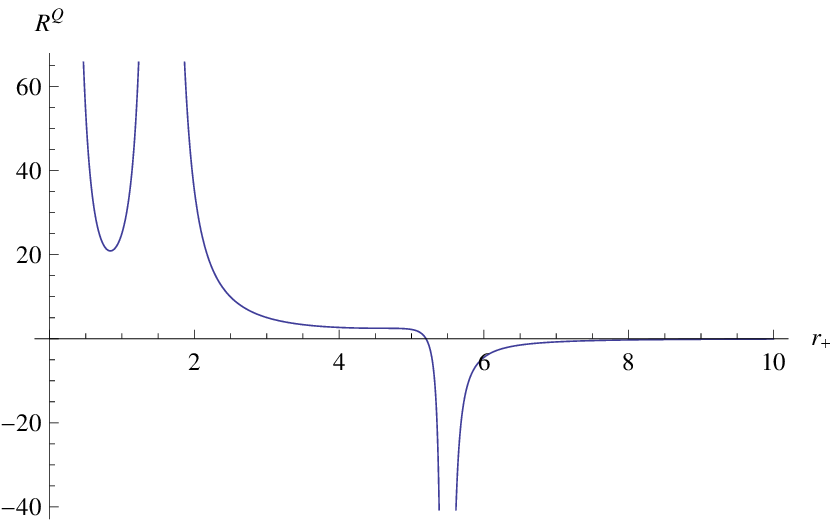}
\figcaption{The Quevedo curvature $R^{Q}$ vs $r_{+}$ with
$b=0.5,l=10,Q=1$. At $r_{+}=0,  1.5307,  5.4781$, $R^{Q}$ diverges.
At $r_{+}$=5.2002/4, it vanishes.} \label{fig11}
\end{center}

\begin{center}
   \includegraphics[width=10cm]{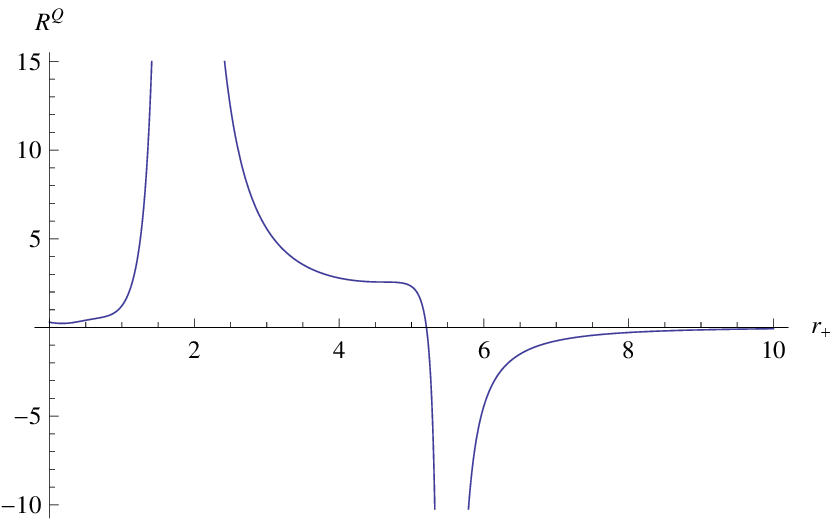}
\figcaption{The Quevedo curvature~$R^{Q}$ vs $r_{+}$ with
$b=1,l=10,Q=1$. At $r_{+}=1.7685/6, 5.4774/5$, $R^{Q}$ diverges. At
$r_{+}$=5.2064/5, it vanishes.} \label{fig12}
\end{center}

Comparing Figure 1, 2 with Figure 11, 12 we can see that the
singular points of the Legendre-invariant metric are consistent with
the result of heat capacity for fixed charge, giving the same
singularities.

\subsection{The free-energy metric}
In~\cite{lvhong}, a new thermodynamical metric was introduced based
on the Hessian matrix of several free energies. The authors
demonstrated that the divergence of the heat capacity corresponds to
the curvature singularities of this new metric. Let's consider the
Helmholtz free energy $F=M-TS$, as a function of $T$ and $Q$, which
satisfies
\begin{equation}
dF=-SdT+\Phi dQ
\end{equation}

The corresponding metric is given by
\begin{equation}
ds^2(F)=-dTdS+d \Phi dQ
\end{equation}
It is not always convenient to use natural variables$(T,Q)$ to
construct the metric, in this case we will use $(r_{+},Q)$
variables. The metric can be obtained straightforward. The Ricci
scalar for the free-energy metric is plotted as follows

\begin{center}
   \includegraphics[width=10cm]{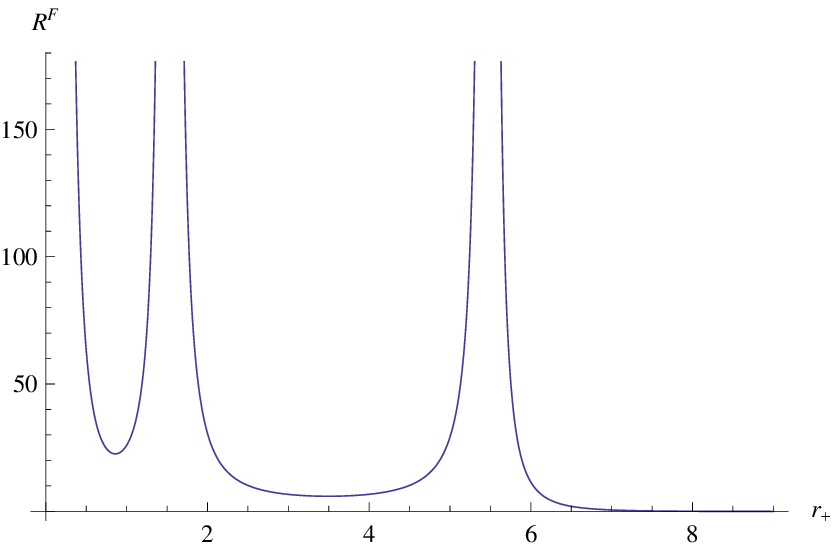}
\figcaption{The curvature obtained using free-energy metric $R^{F}$
vs $r_{+}$ with $b=0.5,l=10,Q=1$. At $r_{+}=0,  1.5307,  5.4781$,
$R^{F}$ diverges. At $r_{+}$=8.1644/5, it vanishes.} \label{fig13}
\end{center}

\begin{center}
   \includegraphics[width=10cm]{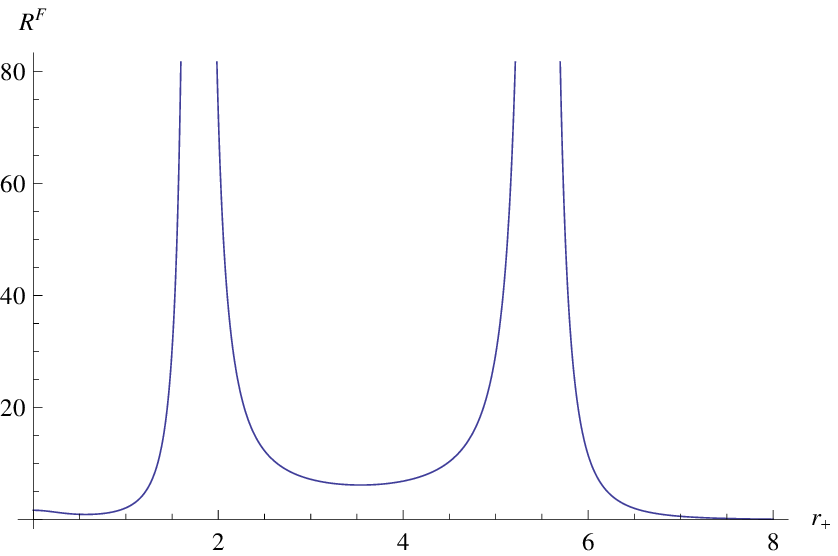}
\figcaption{The curvature obtained using free-energy metric~$R^{F}$
vs $r_{+}$ with $b=1,l=10,Q=1$. At $r_{+}=1.7685/6, 5.4774/5$,
$R^{Q}$ diverges. At $r_{+}$=8.1648/9, it vanishes.} \label{fig14}
\end{center}

Like the Quevedo metric, the free-energy metric reproduces the
singularities of $C_{Q}$. We can also use the potential
\begin{equation}
\bar F=M-\Phi Q
\end{equation}

However, thermodynamical potentials $(U, \bar U)$ are called a
conjugate pair if they satisfy
\begin{equation}
U + \bar U = 2M - TS - \sum_{i=1}^n \mu_i N_i\,.
\end{equation}
Their associated metrics are negative of each other~\cite{lvhong},
{\it i.e.}
\begin{equation}
ds^2(U) = - ds^2 (\bar U)\,.
\end{equation}

So the potential $\bar F$ gives the same information about phase
structure. For the same reason, the potential
\begin{equation}
\bar M=M-TS-\Phi Q
\end{equation}
is conjugate to the potential $M$ (leading to the Weinhold metric),
providing nothing new.
\section{Summary and discussions}

In summary, we have analyzed the thermodynamic geometry of the
4-dimensional BIAdS. The Weinhold, Ruppeiner and a
Legendre-invariant metric are obtained, their scalar curvatures are
computed. The Weinhold geometry is curved for the BIAdS and the
corresponding curvature diverges at some points, but these points
are not the ones at which the heat capacity for fixed charge
diverges or vanishes. The Ruppeiner metric diverges at the same
points as the Weinhold(except two points in the $b=1$ case)and it
gives two more singularites, one of which is the zero point of heat
capacity for fixed charge. For the Legendre-invariant Quevedo metric
and the metric obtained using the free energy as the thermodynamic
potential, they present the same divergent points with heat
capacity. The totality of curvature singularities of all metrics is
exactly the same as the totality of capacity divergent points.

\medskip
\section*{Acknowledgments}
I appreciate the helpful discussions with Ya-peng Hu, Jian-feng Wu,
Hai-qing Zhang, Zhuo-peng Huang and Hai-shan Liu. Special thanks
goes to Prof. Ming Yu, Rong-gen Cai and Hong L\"u for their
warmhearted support and useful suggestions.

\end{document}